\documentclass[journal]{IEEEtran}

\usepackage{caption}
\usepackage{subcaption}
\ifCLASSINFOpdf
  \usepackage[pdftex]{graphicx}
  \graphicspath{{./graph/}}
  \DeclareGraphicsExtensions{.pdf,.jpeg,.png}
\else
  \usepackage[dvips]{graphicx}
  \graphicspath{{./graph/}}
  \DeclareGraphicsExtensions{.eps}
\fi

\ifCLASSINFOpdf
\else
\fi

\usepackage[cmex10]{amsmath,mathtools}
\usepackage{mdwmath}
\usepackage{mdwtab}
\usepackage{amssymb}

\usepackage{algorithmic}
\usepackage{algorithm}
\usepackage{float}

\usepackage{epstopdf}
\usepackage{color}
\usepackage{diagbox}
\usepackage{cite}

\begin{document}

\title{When Internet of Things meets Metaverse: Convergence of Physical and Cyber Worlds}

\author{Kai Li, Yingping Cui, Weicai Li, Tiejun Lv, Xin Yuan, Shenghong Li, \\
Wei Ni, Meryem Simsek, and Falko Dressler
\IEEEcompsocitemizethanks{\IEEEcompsocthanksitem K.~Li is with CISTER Research Centre, Porto 4249-015, Portugal, and also with CyLab Security and Privacy Institute, Carnegie Mellon University, Pittsburgh, PA 15213, USA (E-mail: kai@isep.ipp.pt \& kaili2@andrew.cmu.edu).
\IEEEcompsocthanksitem Y.~Cui, W.~Li, and T.~Lv are with the School of Information and Communication Engineering, Beijing University of Posts and Telecommunications, Beijing 100876, China (E-mail: \{cuiyingping, liweicai, lvtiejun\}@bupt.edu.cn).
\IEEEcompsocthanksitem X.~Yuan, S.~Li, and W.~Ni are with the Digital Productivity and Services Flagship, Commonwealth Scientific and Industrial Research Organization (CSIRO), Sydney, NSW 2122, Australia (E-mail: \{xin.yuan, shenghong.li, wei.ni\}@data61.csiro.au).
\IEEEcompsocthanksitem M.~Simsek is with VMware, Inc., Palo Alto, CA 94304, USA (E-mail: msimsek@vmware.com).
\IEEEcompsocthanksitem F.~Dressler is with the School of Electrical Engineering and Computer Science, TU Berlin, Berlin 10587, Germany (E-mail: dressler@ccs-labs.org).}
}

% The paper headers
\markboth{XXXX,~2022.}%
{Li \MakeLowercase{\textit{et al.}}: }

\IEEEcompsoctitleabstractindextext{%
\begin{abstract}
\boldmath
In recent years, the Internet of Things (IoT) is studied in the context of the Metaverse to provide users immersive cyber-virtual experiences in mixed reality environments. This survey introduces six typical IoT applications in the Metaverse, including collaborative healthcare, education, smart city, entertainment, real estate, and socialization. In the IoT-inspired Metaverse, we also comprehensively survey four pillar technologies that enable augmented reality (AR) and virtual reality (VR), namely, responsible artificial intelligence (AI), high-speed data communications, cost-effective mobile edge computing (MEC), and digital twins. According to the physical-world demands, we outline the current industrial efforts and seven key requirements for building the IoT-inspired Metaverse: immersion, variety, economy, civility, interactivity, authenticity, and independence. In addition, this survey describes the open issues in the IoT-inspired Metaverse, which need to be addressed to eventually achieve the convergence of physical and cyber worlds.
\end{abstract}

}

\maketitle

\IEEEdisplaynotcompsoctitleabstractindextext
\IEEEpeerreviewmaketitle

%=============================================================================%
%============================ Section 1 Introduction================================%
\section{Introduction}
\label{sec_intro}

\subsection{IoT applications in Metaverse} 
Metaverse aims to enable an integrated network of 3D virtual and physical worlds, where a single and universal Internet is extended to provide users immersive cyber-virtual experiences in physical worlds. In particular, two popular applications, i.e., augmented reality (AR) and virtual reality (VR), are developed to bring connected immersive digital experiences and social connections to the Metaverse's users. According to Consumer Technology Association's reports, AR/VR in the digital health care market is estimated to grow from \$960 million in 2019 to \$7 billion in 2026 due to the COVID-19 pandemic~\cite{2022health}. A comprehensive research report by Market Research Future shows that AR/VR in the education market can grow at an 18.2\% compound annual growth rate in the next five years~\cite{2022EducationMarketResearch}.

To support wireless and seamless connected immersive digital experiences, the Internet of Things (IoT) can be leveraged in the Metaverse~\cite{kanter2021metaverse}, which maps real-time IoT data from real life into a digital reality in the virtual world. The IoT can supplement the experiential interface of the users into the virtual world created by AR/VR~\cite{pereira2021arena}. Take the health awareness application as an example. The medical IoT devices can be attached to the user's body or a sensor-laden body suit to instrument the user's state, such as the health conditions that might elicit a response in the virtual realm~\cite{sodhro2018convergence}. The IoT can also help improve e-commerce experiences of the virtual fitting room, where the IoT devices are used to track the movement of the user's body. The personal body information could be updated through data from photos taken on the user's smartphone or smart weighing scales, for instance. This allows the users in the Metaverse to fully immerse in a virtual representation of the store, which overcomes the experiencing barriers of traditional online shopping. In addition, the IoT data can be utilized by the recent Tactile Internet that builds a network or network of networks for remotely accessing or controlling real or virtual objects in real time by humans or machines~\cite{promwongsa2020comprehensive,aijaz2018tactile}. The IoT data can provide context and situational awareness of physical things to AR/VR applications, while triggering data exchange between the digital and physical worlds~\cite{lu2021flash}. For instance, an AR device can react to the user's finger gestures, or trigger a cyber-physical function driven by an event occurring in the physical world. 

The IoT-assisted interaction between real life and the virtual world helps create a digital twin, a digital reflection of the physical state and condition of a unique physical thing~\cite{minerva2020digital}. To achieve a practical digital twin, the Metaverse aims to ensure the reflection is as close to the real-time physical state as possible. Due to this distinct feature, digital twins grow as one of the fundamental applications in the Metaverse. In professional settings, digital twins can be constructed with the Tactile Internet as well as the Haptic Codecs (IEEE P1918.1.1)~\cite{steinbach2018haptic} to make a group meeting productive since the users can interact with each other while operating or demonstrating a replica of the hardware or software prototype. In technical training programs, digital twins help engineers directly operate 3D representations of complex systems~\cite{stojanovic2018data}. The digital twins can also recreate a complete maintenance workshop with a virtual copy of serviced devices and some mechanical tools for repairing the device. The digital twins set up a virtual environment connected with a real workshop in the physical world, where the maintenance can be carried out remotely. In urban planning and construction, the digital twins can virtualize a real-world city, where citizens or economic players can use visual representation to implement a development plan and discover future urban projects~\cite{farsi2020digital}.

\begin{figure*}[htp]
	\centering{}\includegraphics[scale=0.6]{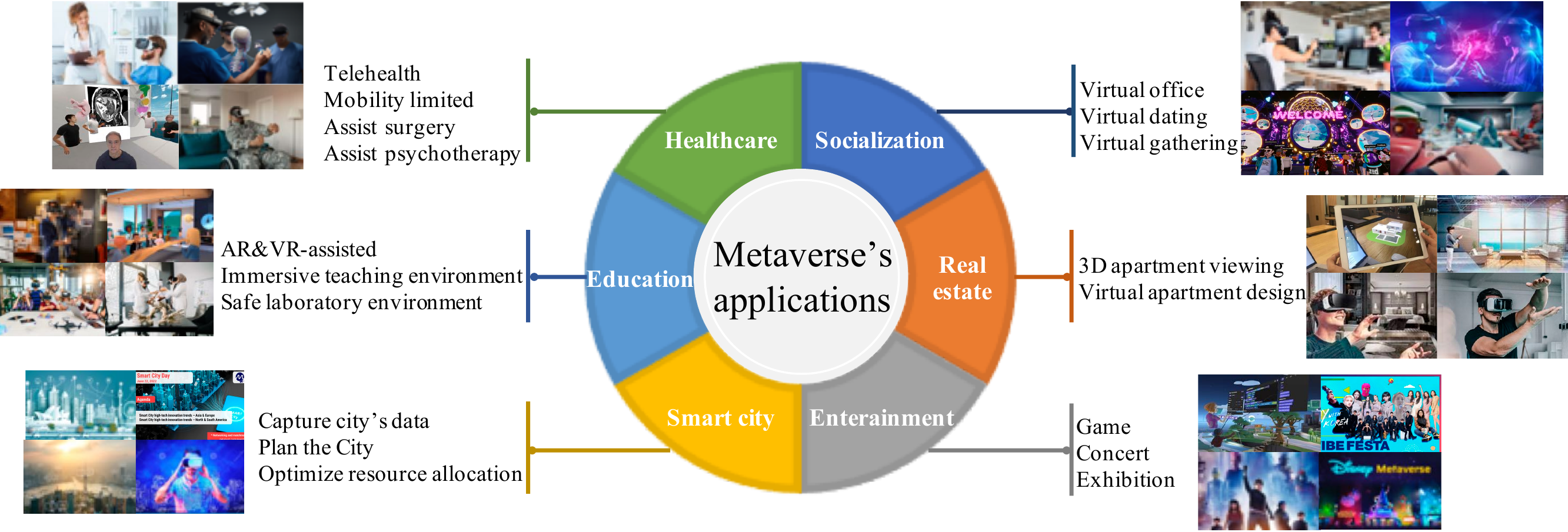}
	\caption{The six typical applications of the Metaverse include healthcare, education, smart city, entertainment, real estate, and socialization.}
	\label{fig:app}
\end{figure*}

The Metaverse can be applied to several transaction scenarios in the digital economy application while achieving a high-efficiency trade. Essentially, Figure~\ref{fig:app} introduces the typical 6 application scenarios of the Metaverse, including collaborative healthcare~\cite{thomason2021metahealth}, education~\cite{kye2021educational}, smart city~\cite{8710517}, entertainment~\cite{lee2021study}, real estate~\cite{nalbant2021computer}, and socialization~\cite{ning2021survey}.

\textit{Healthcare}: We are hurtling into the age of the metaverse. The World Economic Forum has predicted that the introduction of digital services is expected to be one of the most critical factors in transforming healthcare over the next decade~\cite{thomason2021metahealth}. Traditional healthcare is a hands-on personal encounter where doctors and patients meet face-to-face for diagnosis and treatment. The Covid-19 pandemic~\cite{maatuk2022covid} has forced people to look for alternatives that enable patients to be managed out of the hospital remotely. Before the pandemic, 43\% of healthcare facilities could provide telehealth, which rose to 95\% in 2020~\cite{demeke2021trends}. The Metaverse promises to facilitate the development of telehealth, the patients can immerse themselves in the Metaverse through their avatar to have a health consultation with their doctors.

The Metaverse can help people with disabilities, or limited mobility live better. For example, people with hearing or visual weakness can benefit from the Metaverse technology (e.g., AR or VR) to communicate better; people with paraplegia can enjoy a healthy life in the Metaverse. 
In addition, AR becomes a promising application to empower the skills and knowledge-based medical students. For instance, many technology companies, such as Microsoft Hololens~\cite{hanna2018augmented}, are developing surgical assistive tools for on-demand surgical operations. Moreover, the Metaverse can help with psychotherapy~\cite{ning2021survey}. A virtual and relaxing environment can be constructed through the Metaverse in which people with mental disorders can communicate and interact with avatars.

\textit{Education}: In recent years, the Metaverse technologies in the education sector have gained significant momentum~\cite{kye2021educational}. AR/VR that emphasizes visualization-based learning concepts is widely used to transform traditional teaching approaches~\cite{rajagopal2019improving}. For creating engaging and immersive learning environments for students, the Metaverse is expected to extend a variety of learning institutions, which can visualize the teaching content and promote understanding of the learning content. The Metaverse can also create virtual, safe, but immersive lab environments. 
Moreover, some learning materials are difficult to observe directly or explain in text~\cite{hans2020researchAR}, e.g., the human body's structure and functions of an organ system, or the universe. To enhance the understandability of the learning materials, AR/VR can be developed to effectively assist the students with the required constant practice and experience. For example, Curiscope, an AR company based in Brno, Czech Republic, developed Virtuali-Tee, an AR-enabled T-shirt~\cite{Virtuali_Tee}. Virtuali-Tee allows the students to visualize and examine the inside of the human body as if they were in an anatomy lab.

\textit{Smart city}: As a supporting technology in the Metaverse, IoT-empowered digital twins can utilize the IoT data to digitize the objects in the physical world, such as roads and streets, houses, vehicles, and city infrastructures, and create virtual cities. This is a useful tool for building smart cities~\cite{8710517}. With the digital twins, developers and constructors can easily build a simulation environment that comprehensively mirrors the physical world. The digital twins can capture the elements in the smart city, such as people, vehicles, traffic lights, and buildings. The IoT data helps map the physical world to form a visible, controllable, and manageable digital twins city. Essentially, the Metaverse is used in the smart city to optimize the allocation of facilities and resources. For example, Intel and Siemens in Germany collaboratively develop a smart parking solution, where the digital twins-assisted meta-boundary platform is used to manage the local road traffic~\cite{chang20226g}.

\textit{Entertainment}: The development of Metaverse-related technologies has greatly improved the immersion of games, which can effectively enhance user experience, playability, and fun. A representative Metaverse game is Roblox relying mainly on VR technology, which has 150 million monthly active users~\cite{meier2020using}. Along with the widespread use of payments and personal information in the Metaverse, a game based on blockchain technology has been proposed in~\cite{3174952}. The Metaverse also provides an immersive environment and can be used for large events such as exhibitions, concerts and book signings. For example, the Korean pop artists Asepa and Black Pink released new songs and held fan signings in the virtual world built by the Metaverse~\cite{lee2021study}.

\textit{Real estate}: This has been listed as the next promising application entry in the Metaverse~\cite{nalbant2021computer}.
As one of the key technologies driving the Metaverse, AR/VR provides customers with a realistic and immersive experience~\cite{hou2020technology}. For example, real estate agents can harness the power of VR to provide buyers with an immersive virtual tour of a property. This offers the customer the advantage of conducting inspections in all conditions and reduces the time spent out of the house to zero.

Additionally, with the help of AR/VR and AI technology, the Metaverse has a strong potential to deploy in the interior design and architectural fields since it can provide an immersive virtual world and enable real-time interaction~\cite{9734970}. We can not only experience the full spherical panoramic view of the interior and exterior of the house, but also directly change the furniture and internal architectures, and see our dream home in real-time. For example, Beike offers a virtual renovation platform~\cite{zhang2021leading} where customers can see what they imagine their renovated home will look like, which helps consumers see the potential of the house. Urbanbase has developed ``AR Scale'', a 3D cloud-based augmented reality presentation service~\cite{kim2021bim}. Architectural professionals do not need to create real models to show others actual building design solutions. Instead, they just display 3D models on the actual building site through AR scale using a 1:1 scale model to check harmony in advance with the surrounding environment and buildings.

\textit{Socialization}: The Metaverse opens a new application field in terms of human social forms~\cite{ning2021survey}. The Metaverse breaks the boundaries of time and space, which can provide various forms of social interaction, bringing people closer to each other~\cite{kim2021study}. People can pursue some higher-level needs beyond the level of the physical world, such as virtual offices, virtual dating, and virtual gatherings. With the impact of the COVID-19 pandemic, the importance of telecommuting and remote social networking has become apparent~\cite{lyttelton2022telecommuting}. The Metaverse can compensate for the limitations of traditional models and further improve the functionality of telecommuting and remote social platforms to create a more authentic social environment.

\subsection{Key contributions of this survey} 
In this paper, we comprehensively survey four pillar technologies with their typical applications in the IoT-empowered Metaverse, namely, responsible artificial intelligence (AI), high-speed data communications, and cost-effective mobile edge computing (MEC), and digital twins. 

Since building the Metaverse is highly complex, the learning process is turned into what is commonly referred to as a ``black box'', which is difficult to interpret. The responsible AI can be applied to perceive how the IoT functions in the Metaverse. Specifically, the responsible AI is a technique that helps the engineers or data scientists who develop AI algorithms explain what is exactly happening inside the IoT applications and how the AI algorithm leads to a specific result in the Metaverse. Therefore, the AI decision-making processes can be fully understood according to the ethical, moral, legal, cultural, and socio-economic consequences in the Metaverse. 

Providing the connected AR/VR immersive experiences in the Metaverse relies on enormous amounts of data exchange with ultra-low latency, where the nature of haptic signals and human perception requires 1 ms of latency at maximum. Since the bandwidth of 4G-based wireless networks is scarce, the majority of 4G-based AR/VR applications cannot undertake real-time data communication with a large number of IoT devices. For example, LTE wireless systems can only allow a latency of 25 ms at the minimum~\cite{sukhmani2018edge}. Therefore, lots of research focuses on IoT networks with ultra-high-speed data communications for the AR/VR experience.

Keeping the Metaverse's users engaged and immersive is important to provide the same experiences as reality. Essentially, the latency of the IoT device has to be maintained lower than the human perceptible limit. To reduce the latency, the Metaverse applies MEC as a support, where MEC allows the IoT devices to utilize the nearby communication infrastructures and computing resources~\cite{Lyu2017Optimal,cui2019stochastic,lyu2018energy}. By exploiting the MEC, the Metaverse can offer ultra-low-latency immersive experiences to satisfy the human perceptible limit, while powering the virtual 3D realm and IoT networks. MEC can also enable the Metaverse a cross-cloud computing that creates an open environment across multi-vendor cloud platforms~\cite{ahmed2019trust,elkhatib2016mapping}. This benefits the Metaverse's users with on-demand scalability and enhanced service quality. 

As one of the Metaverse's core building blocks, the digital twins create digital replicas corresponding to products and services in the physical world. The digital twins can also visualize data analysis and allow business organizations to carry out simulations based on the physical world's conditions before making costly decisions~\cite{noauthor_digital_nodate, 9103025}. 
Implementing the digital twins requires IoT applications to incorporate low-latency data collection, precise data modeling, and fusion of heterogeneous sensory data, which is challenging. Moreover, the implementation is throughout the product development life cycle, from designing to post-production monitoring and service, to improve product productivity and profitability. 
With the connection of sensing networks and two-way flow of information~\cite{roy_digital_2020}, this technology reflects any changes in the physical world to synchronize the digital environment and the physical world. The digital twins are inter-dependent computer programs that integrate the IoT~\cite{9120192,9320532}, AI~\cite{LI2022167,9198575} and semantic communication~\cite{9380190,austin_architecting_2020} to create simulations and predictions of the performance of products or processes.

Digital twins, which include replicas of building facilities\cite{mohammadi_knowledge_2020}, operational processes~\cite{noauthor_digital_nodate,roy_digital_2020,9198575}, human-computer interaction processes~\cite{malik_digital_2018}, and social services such as healthcare~\cite{9320532} and education~\cite{LiljaniemiPaavilainen}, provide an entry point for the Metaverse to connect to the physical world. By deploying a collection of digital twins, the Metaverse infrastructure is enhanced, and all the potential Metaverse applications are easier to be driven into reality.

The rest of this survey is organized as follows. In Section~\ref{sec_surveys}, we discuss the existing surveys and tutorials on the Metaverse. Section~\ref{sec_tech} describes current industrial efforts and the key requirements for enabling IoT with the Metaverse in daily life. In Section~\ref{sec_pillars}, we investigate the four pillar technologies that support the fundamental applications of the Metaverse: responsible AI, high-speed data communications, cost-effective mobile edge computing, and IoT-empowered digital twins. In Section~\ref{sec_cha}, we outline the open scopes and remaining challenges of applying the four pillar technologies to the AR/VR applications in the Metaverse. Finally, Section~\ref{sec_cond} concludes the survey.

%=============================================================================%
%============================Section 2 Related Work===============================%
\section{Existing Surveys and Tutorials}
\label{sec_surveys}
\begin{table*}[]
\centering
\caption{Existing surveys and tutorials on the Metaverse}
\begin{tabular}{llll}
\centering
\textbf{Paper} & \textbf{Short description} & \textbf{Key technologies} & \textbf{Applications covered}   \\ \hline

\begin{tabular}[c]{@{}l@{}}\cite{park2022metaverse} \\ \textbf{2022}\end{tabular} 
& \begin{tabular}[c]{@{}l@{}}A taxonomy aims to classify the \\ AI technologies in the Metaverse \end{tabular} 
& \begin{tabular}[c]{@{}l@{}}Hardware, software, contents, \\ user interaction, and implementations \end{tabular} 
& \begin{tabular}[c]{@{}l@{}}Ready Player One, Roblox,\\ and Facebook Research \end{tabular} \\ \hline

\begin{tabular}[c]{@{}l@{}}\cite{ning2021survey} \\ \textbf{2021}\end{tabular} 
& \begin{tabular}[c]{@{}l@{}}The survey about the development status\\ of the Metaverse applications \end{tabular} 
& \begin{tabular}[c]{@{}l@{}}Network infrastructure, resource management,\\ connected AR/VR, and blockchain \end{tabular} 
& \begin{tabular}[c]{@{}l@{}}AR/VR, cultural civilization and \\ legal systems, and hyperspace \end{tabular} \\ \hline

\begin{tabular}[c]{@{}l@{}}\cite{lee2021all} \\ \textbf{2021}\end{tabular} 
& \begin{tabular}[c]{@{}l@{}}A digital twins-native continuum is presented \\ to summarize the development of the Metaverse \end{tabular} 
& \begin{tabular}[c]{@{}l@{}}AR/VR, human-machine interaction, AI, \\ robotics, blockchain, and MEC \end{tabular} 
& \begin{tabular}[c]{@{}l@{}}content creation, data interoperability, \\ social acceptability, and security \end{tabular} \\ \hline

\begin{tabular}[c]{@{}l@{}}\cite{wang2022survey} \\ \textbf{2022}\end{tabular} 
& \begin{tabular}[c]{@{}l@{}}A survey of the Metaverse architecture, \\ as well as the security and privacy protection \end{tabular} 
& \begin{tabular}[c]{@{}l@{}}Digital twins, native content creation, \\ and the integration of the physical-virtual reality \end{tabular} 
& \begin{tabular}[c]{@{}l@{}}Identity and data management, \\ digital footprints, and content control \end{tabular} \\ \hline

\begin{tabular}[c]{@{}l@{}}\cite{huynh2022artificial} \\ \textbf{2022}\end{tabular} 
& \begin{tabular}[c]{@{}l@{}}The AI-based frameworks and their \\ functionality are reviewed \end{tabular} 
& \begin{tabular}[c]{@{}l@{}}Natural language processing, machine vision, \\ blockchain, digital twin, and neural interface \end{tabular} 
& \begin{tabular}[c]{@{}l@{}}healthcare, manufacturing, \\ smart cities, and gaming \end{tabular} \\ \hline

\begin{tabular}[c]{@{}l@{}}\cite{wu2020towards} \\ \textbf{2020}\end{tabular}  
& \begin{tabular}[c]{@{}l@{}}The AI development is introduced as a \\ national strategy for technological promotion \end{tabular} 
& \begin{tabular}[c]{@{}l@{}}Advanced driver assistance, deep learning, \\ big data, AI ecosystem \end{tabular} 
& \begin{tabular}[c]{@{}l@{}}Autonomous driving, IoT, \\ medical industry, everyday consumers \end{tabular} \\ \hline

\begin{tabular}[c]{@{}l@{}}\cite{fernandez2022life} \\ \textbf{2022}\end{tabular}  
& \begin{tabular}[c]{@{}l@{}}An overview of the Metaverse development \\ in terms of privacy, governance, and ethics \end{tabular} 
& \begin{tabular}[c]{@{}l@{}}Sensory-level privacy, social interaction \\ privacy, modular governance and ethics \end{tabular} 
& \begin{tabular}[c]{@{}l@{}}AR/VR, head-mounted displays, \\ online games, and social networks \end{tabular} \\ \hline

\begin{tabular}[c]{@{}l@{}}\cite{dionisio20133d} \\ \textbf{2013}\end{tabular} 
& \begin{tabular}[c]{@{}l@{}}The survey summarizes the four features \\ according to the physical-virtual interaction \end{tabular} 
& \begin{tabular}[c]{@{}l@{}}Visual immersion, digital sound, \\ activity and interaction identification \end{tabular} 
& \begin{tabular}[c]{@{}l@{}}Stereoscopic vision, gesture recognition, \\ and virtual world audio environment \end{tabular} \\ \hline
                     
\begin{tabular}[c]{@{}l@{}}\cite{ynag2022fusing} \\ \textbf{2022}\end{tabular} 
& \begin{tabular}[c]{@{}l@{}}Fusing blockchain and AI for the Metaverse \\ users to enjoy the verifiable services \end{tabular} 
& \begin{tabular}[c]{@{}l@{}}Data verification and consensus, \\ distributed storage, and smart contract \end{tabular} 
& \begin{tabular}[c]{@{}l@{}}Digital creation, digital asset, \\ digital market, and digital currency \end{tabular} \\ \hline

\begin{tabular}[c]{@{}l@{}}\cite{mozumder2022overview} \\ \textbf{2022}\end{tabular} 
& \begin{tabular}[c]{@{}l@{}}Blockchain and AI can be applied to \\ IoT-based healthcare applications \end{tabular} 
& \begin{tabular}[c]{@{}l@{}}Communication computing infrastructure, \\ decentralized storage and computation \end{tabular} 
& \begin{tabular}[c]{@{}l@{}}Remote surgery, AR surgery, \\ healthcare facilities, medical therapy \end{tabular} \\ \hline

\begin{tabular}[c]{@{}l@{}}\cite{chang20226g} \\ \textbf{2022}\end{tabular} 
& \begin{tabular}[c]{@{}l@{}}An integration of 6G-assisted \\ edge AI and the Metaverse \end{tabular} 
& \begin{tabular}[c]{@{}l@{}}6G, edge AI, network architecture, \\ resource allocation, and blockchain \end{tabular} 
& \begin{tabular}[c]{@{}l@{}}Immersion education, telecommuting \\ videoconferencing, and production \end{tabular} \\ \hline

\begin{tabular}[c]{@{}l@{}}\textbf{This survey} \\ \textbf{2022}\end{tabular} 
& \begin{tabular}[c]{@{}l@{}}Four pillar technologies that leverage IoT networks \\ for an integration of the physical and cyber world. \\ The challenges and open issues of the future \\ development of the IoT-empowered Metaverse. \end{tabular} 
& \begin{tabular}[c]{@{}l@{}}Responsible AI, RIS, mmWave, NOMA, \\ FL, edge AI, and IoT with digital twins \end{tabular} 
& \begin{tabular}[c]{@{}l@{}}Smart healthcare, industry 4.0, smart \\ agriculture, data visualization, , \end{tabular} \\ \hline

\end{tabular}
\label{tb_surveys}
\end{table*}

In this section, we review some recent studies on the applications of AI, MEC, and virtual systems in the Metaverse.

\subsection{Recent studies}
The Metaverse, integrating social activities and AI methods, incubates service demands from mobile-based always-on access to connectivity with reality using virtual currency. A tutorial in~\cite{park2022metaverse} presents a Metaverse taxonomy, which classifies the technologies based on hardware, software, contents, user interaction, implementations, and applications. The taxonomy is applied to three case studies on Ready Player One, Roblox, and Facebook Research, where the social influence and technical limitations of the Metaverse are discussed. 
The authors in~\cite{ning2021survey} survey the Metaverse's development in terms of network infrastructure, resource management, connected VR, and its convergence. Three characteristics of the Metaverse are introduced, namely, multi-technology dominance (including AR and blockchain-based economic systems), sociality (including cultural and legal systems), and hyperspace (which breaks the boundary between the virtual world and the real world). 
A digital twins-native continuum is presented in~\cite{lee2021all}, which summarizes the development of the Metaverse as three steps, i.e., digital twins of humans and IoT, native content creation, and integration of the physical and virtual worlds. Several technologies, e.g., AR/VR, human-machine interaction, AI, robotics, blockchain, and MEC, are explored to build the ecosystems in the Metaverse, which can be applied to the application of content creation, data interoperability, social acceptability, and security. 
The survey~\cite{wang2022survey} focuses on the security and privacy issues of the Metaverse. Similar to~\cite{lee2021all}, the development of the Metaverse contains three steps, including digital twins, native content creation, and the integration of the physical-virtual reality. In an AI-enabled virtual scene, AR/VR devices equipped with IoT sensors can collect brain-wave signals, facial expressions, eye or hand movements, and environmental conditions~\cite{sharma2020all}. Adversaries can invade and control the AR/VR devices to track the users' locations, biometric features, and users' identities, which threatens human safety and critical IoT infrastructures. Current security protection solutions for the Metaverse, e.g., identity and data management, avatar clone, digital footprints, situational awareness, and user-generated content control, are discussed with regard to the application requirement. 

In~\cite{huynh2022artificial}, the authors review the AI-based frameworks, e.g., supervised learning, unsupervised learning, and reinforcement learning, as well as their functionality in the Metaverse. Several popular AI techniques, such as natural language processing, computer vision, blockchain, networking, digital twin, and neural interface, are also studied to improve the user's virtual experience in physical applications, such as health monitoring, intelligent transportation, industrial production, and online shopping. 
In~\cite{wu2020towards}, future AI development is outlined at a national level, which can redefine the economic activities, production, and business demand in a country. The authors take China as an example and list the main national AI structures, including National Open Innovation Platforms, National NGAI Development Experimental Zones, and Pilot zones for innovative application of AI. Most of China's information technology companies are involved in the structures to provide AI solutions for online consumer credit services, autonomous driving, clinical workflow, and customizable virtual conversation assistants. 

The Metaverse applications, e.g., AR/VR, and head-mounted displays, capture a vast amount of biometrical data, threatening the users' privacy and security. Considering humanity, equality, and diversity, the monitoring tools should be developed to deal with the user's misbehavior in the Metaverse, while the tools also encourage positive behaviors. Thus, the tutorial in~\cite{fernandez2022life} envisions that the application design in the Metaverse is guided following three aspects: privacy, governance, and ethical consideration.
The development of virtual worlds in the past decades, from the individual text-based environment to the Metaverse, is presented in~\cite{dionisio20133d}. The survey summarizes four features according to the physical and virtual interaction: realism (enhancing the immersed experience for the user), ubiquity (allowing the full system access for the existing IoT devices while maintaining the user's virtual identity over different applications), interoperability (enabling the user to have a seamless uninterrupted virtual experience), and scalability (supporting a large number of users to run the Metaverse applications concurrently).

In~\cite{ynag2022fusing}, the authors present how blockchain and AI can be fused into an economic system composed of digital creation, digital assets, digital market, and digital currency. Due to enhanced traceability and confidentiality of the data, the users in the Metaverse can enjoy immutable and verifiable services, such as cryptocurrency exchanges, and low-latency transactions and authentication. 
Moreover, blockchain and AI can be applied to IoT-based healthcare applications~\cite{mozumder2022overview}. The users' medical data are stored at different servers worldwide, where the Metaverse processes personal data using AI techniques. The decentralized storage and computation can extend the scalability of the services and prevent malicious attacks. 

The survey~\cite{chang20226g} discusses 6G-assisted AI techniques for the real-time interaction of the users. A basic architecture of the Metaverse is studied with a physical layer, a virtual layer, and a technical layer. For serving large-scale AI models, three 6G-enabled architectures, i.e., edge cloud-Metaverse, mobile edge cloud-Metaverse, and decentralized Metaverse, are presented to improve resource allocation efficiency and computing capabilities. 

\subsection{Our Contributions}
The existing surveys and tutorials describe the Metaverse structure, AR/VR applications, digital twins, AI, and blockchain technologies. In contrast, our survey specifically focuses on the IoT-empowered Metaverse, where the real-time IoT data are leveraged to map the physical world into a digital reality in the virtual world. We provide a fine-grained study of the four pillar technologies that support fundamental IoT applications in the Metaverse, including responsible AI, high-speed data communications, cost-effective MEC, and IoT-empowered digital twins. Furthermore, we expatiate the challenges and open issues of the future development of the IoT-empowered Metaverse, such as data processing, security and privacy, real-time 3D modeling, scalable cyber worlds, connected mobile users' experience, interoperability and uniformity of virtual platforms, and the barriers of the physical world (details refer to Section~\ref{sec_cha}).

%=============================================================================%
%============================Section 3 metaverse requirements==========================%
\section{Industrial Efforts and Key Requirements For Enabling IoT with Metaverse in Daily Life}
\label{sec_tech}

{
The Metaverse is a massively scaled, interoperable platform for virtual activities, spatially merges the physical with the virtual world, but in perceived real-time~\cite{encyclopedia2010031}. It can provide an experience that spans both the digital and physical worlds, private and public networks/experiences, and open and closed platforms. In the Metaverse, an unlimited number of participants with with a sense of individual presence can synchronously and persistently enjoy the same or even beyond reality type of experiences they would in the physical world~\cite{lee2021all}. Each participant corresponds to an identity of the Metaverse; and can drop into the Metaverse to conduct social and spiritual activities, e.g., healthcare and education~\cite{thomason2021metahealth}, office~\cite{ning2021survey}, social, trading, creation, and entertainment~\cite{ynag2022fusing} as needed, and then pop back out to reality. All these event spaces are anticipated to be interoperable, and their data is continuous. The Metaverse will bring a new level of consistency and immersion to today's increasingly popular hybrid event formats.

\newcommand{\tabincell}[2]{\begin{tabular}{@{}#1@{}}#2\end{tabular}}  
\begin{table}[ht]
\centering
\caption{The representative companies and their typical products}
\label{ta:company}
\begin{tabular}{|p{2.3cm}|p{5.2cm}|}
\hline  
\textbf{Game maker}&Roblox (Roblox);\newline Tencent (Roblox Chinese ver.);\newline Meta (Horizon World);\newline Amazon (AMS Cloud Quest);\newline Epic Games (Fortnite); \newline ByteDance (Restart the World).\\
\hline  
\textbf{Live entertainment}& Disney (Theme park Metaverse); \newline Avex Business Development and \newline Digital Motion (Virtual Avex Group).\\
\hline
\textbf{Social platform}& ByteDance (Pixsoul, Party Island);\newline Baidu (Land of Hope);\newline Meta (Facebook Horizon).\\
\hline 
\textbf{Office}& Microsoft (Mesh for Teams);\newline Accenture (One Accenture Park).\\
\hline
\textbf{Hardware}&Meta (Oculus Quest);\newline Samsung (Gear VR); \newline Google (Google Cardboard);\newline Sony (PSVR2); \newline Lenovo (ThinkReality).\\
\hline
\textbf{Platform}& Nvidia (Omniverse);\newline Sony and Hassilas (Mechaverse).\\
\hline 
\textbf{Algorithm design \newline and optimization}& Google (DeepMind team);\newline OpenAI lab;\newline Baidu (Institute of Deep Learning).\\
\hline 
\end{tabular}
\end{table}

With the COVID-19 pandemic, people have to stay home for many activities, which has led an increasing number of people to rely on media and technology to inform and thus entertain, educate, and socialize~\cite{zallio2022inclusive}. Consequently, the Metaverse is realized as the next disruptive technology~\cite{huynh2022artificial} and currently attracting enormous attention from around the world, since it enables novel forms of engrossing tele-presence~\cite{duan2021metaverse}.

Many companies are dedicated to launching various metadata software types, covering areas such as gaming, video conferencing, game development, and AI face painting.
In Dec. 2021, Meta released the game ``Horizon World''~\cite{kraus2022facebook}. Participants can use virtual reality (VR) headsets to create their own avatars and interact with other participants' avatars in this virtual world.
ByteDance is now testing its first Metaverse social app, called ``Party Island'', which creates a parallel online universe where people can communicate and join events in real-time through avatars.
In addition, some companies work on hardware and operating systems related to the Metaverse, with VR, AR, and AI the most, and the application scenarios are mostly about games and social~\cite{9728808}. There are also companies laying out Metaverse-related technologies in other areas. For example, Lenovo proposed the concept of ``HoloBoard''~\cite{3474761}, an immersive future blackboard that uses holographic projection to create a more immersive teaching classroom.
Moreover, some companies have started to make efforts in building and optimize the corresponding technologies, including the underlying architecture of the Metaverse (Amazon and Roblox)~\cite{3479238, wang2022survey}, compute capability, and algorithm optimization (Google, Baidu, and Microsoft)~\cite{lewis2021vogue,Xu_2021_CVPR, liu2021swin}.
The representative companies and their products in the area of Metaverse are shown in Table~\ref{ta:company}.

The IoT-empowered Metaverse has seven key requirements, which are immersion, variety, economy, civility, interactivity, and authenticity and independence, as listed in Table~\ref{ta:meta_fea} and described in the following. Accordingly, those requirements demand further development of IoT technologies. 
\begin{table*}[ht]
\centering
\caption{The typical requirements of the Metaverse and technical demands of IoT}
\label{ta:meta_fea}
\begin{tabular}{|p{7.15cm}|p{10.1cm}|}
\hline 
\textbf{Requirements of the Metaverse} & \textbf{Technical demands of IoT}\\
\hline  
Immersion \newline(sustainable, synchronous and live) & High synchronization, low latency, huge data monitoring, and real-time computing; \newline Ensure network security, privacy, and trust.\\
\hline  
Anywhere, anytime and any participant& The limitation of the network;\newline The barrier of the platforms;\newline Ensure network security, privacy, and trustworthiness;\newline The contradiction between light-weight and computing power, electricity, and diversification;\newline The need of resources to make this virtual world work.\\
\hline  
Variety \newline(heterogeneity,\,diverse\,events, places and activities) &  Ensure interoperability.\\
\hline  
Economy \newline(fully functioning and independent creator economy) & High synchronization, low latency, huge data monitoring, and real-time computing;\newline
Ensure network security, privacy, and trust.\\
\hline  
Civility \newline(diversity, equality, and inclusiveness)& Different languages and civilizations, how to communicate? Can we rely on SemCom? \newline
Local semantic knowledge base (Sem-KB) vs. cloud Sem-KB.\\
\hline  
Interactivity \newline(seamless connection) & High synchronization, low latency, huge data monitoring and real-time computing; \newline
The barrier of the platform; \newline
Ensure interoperability.\\
\hline  
Authenticity\, and\, independence \newline(digital copies of the physical world and parallel space) & The accuracy of introducing the physical world into the Metaverse.\\
\hline
\end{tabular}
\end{table*}

\subsection{Immersive Experience}
For the participants to experience the Metaverse as they would experience the physical world, the nature of high immersion is essential to the Metaverse~\cite{mystakidis2022metaverse}.
The Metaverse can be experienced synchronously and persistently by an unlimited number of participants with an individual sense of presence, and with continuity of data, such as identity, entitlements,  objects, history, communications, and payments~\cite{david2022nft}.
Events of any size can be hosted in the Metaverse, such as major trade fairs, exhibitions, etc. 
Moreover, the Haptic Codecs and Tactile Internet can be developed in the Metaverse to exchange multi-modal data, such as the combination of audio, video and haptic information, over the Internet. This will enable the users experience with an enhanced virtual presence, immersing in a remote environment~\cite{wang2022towards,strese2019haptic}. 

\subsection{Anywhere, Anytime and Any participant}
The Metaverse is a sustainable, widely covered virtual world, and user access to the Metaverse is not limited by location or time~\cite{alpala2022smart}.
Anywhere has two main meanings, a) Anywhere in the Metaverse: The Metaverse can model any scenario and guarantee that the participants successfully access the location or scenario they want to visit; b) The participants accessing the Metaverse are not limited by the location: The Metaverse is decentralized, and each participant can access the Metaverse from anywhere with a simple device.
Anytime: The participants can access the Metaverse at any time, since time in the Metaverse is continuous.
Any participant: There is no limit to the identity and number of concurrent participants in the Metaverse.

\subsection{Variety of Events, Places, and Activities}
There will be various events and activities happening in diverse places in the Metaverse at all times. 
Everyone can be a part of the Metaverse and participate in specific activities and events together and simultaneously with other participants.
To achieve freedom and diversity beyond reality, including a feature that cannot be ignored: User Generated Content~\cite{3479238}, the Metaverse should also provide each participant with an individual sense of ``presence''.

\subsection{Fully Functioning Economy} 
Unlike the traditional economy, the Metaverse has an independent and fully functioning creator economy within the system~\cite{seigneur2022should}.
Participants can create and trade new assets or experiences in the virtual world created by the Metaverse while enjoying full ownership~\cite{kaur2021metaverse}. Creators can use their assets or experiences in the Metaverse and trade them for desired values.
Individuals and businesses can create, own, invest in, and sell a wide variety of ``works" and be rewarded for the ``value" they generate that others recognize.

\subsection{Independent and Complete Civilization System}
The Metaverse has its own system of civilization~\cite{bowen2022metaverse}. We have a life in it, a few people may form communities, and communities make up large cities, recognize villages, cities, and even all kinds of rules.
Like a social circle, the participants with different languages and cultural backgrounds drive various cultural clusters in the Metaverse.
We make the rules in the Metaverse, follow them, and then live here together to evolve into a civilized society.
everyone makes common rules and then lives on together in it, evolving into a civilized society.

\subsection{Interactivity}
The Metaverse enables seamless connectivity between users and users, between users and platforms, between platforms and platforms, and between operating systems and operating systems, etc~\cite{wang2022survey,mystakidis2022metaverse}. We can create and use assets, avatars, and experiences in any accessible Metaverse without restrictions. In the Metaverse, data, digital items/assets, content, etc. can be interoperable in unprecedented ways. For example, skins of our guns in Counter-Strike can also be used to decorate guns in Crossfire or gifted to friends via Facebook.

\subsection{Authenticity and Independence}
There are digital copies of the physical world and creations of the virtual world in the Metaverse~\cite{hollensen2022metaverse}.
It is a parallel space closely connected to the external reality and highly independent. 
Participants in the Metaverse can not only do what they can do in the physical world but also achieve things that transcend reality in space.
}

%=============================================================================%
%==========================Section 4 Four Pillar Technologies=========================%
\section{Four Pillar Technologies of Metaverse}
\label{sec_pillars}
In this section, we present the four pillar technologies that support fundamental applications of the Metaverse, namely, responsible AI, high-speed data communications, cost-effective mobile edge computing, and IoT-based digital twins. 

\begin{figure*}[htb]
\begin{center}
\includegraphics[width=0.7\textwidth]{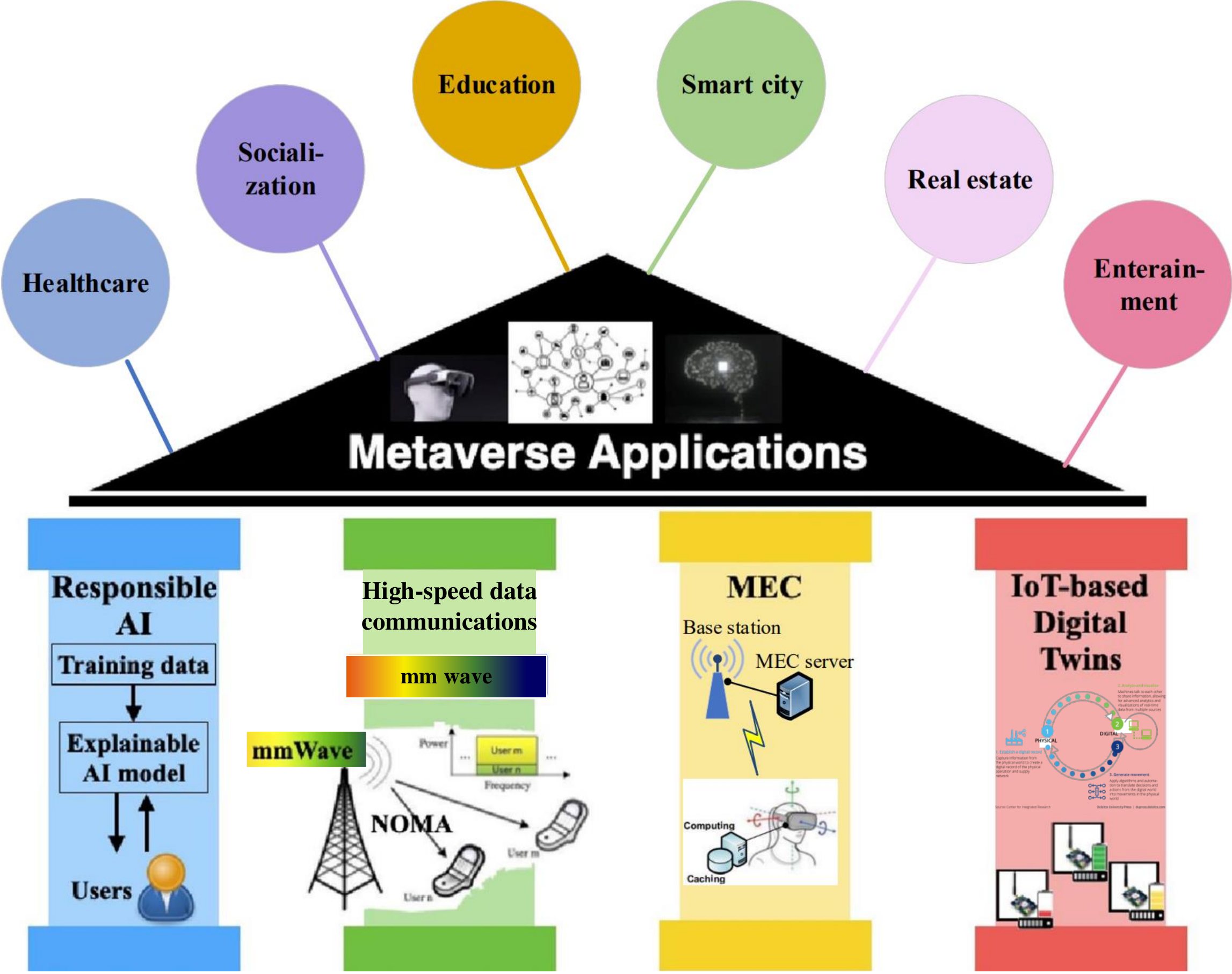}
\end{center}
\caption{The four pillar technologies that support the IoT applications in the Metaverse. The responsible AI helps engineers or data scientists explain what exactly is happening inside the AI algorithm; the high-speed data communication empowers real-time connections based on several promising technologies, such as millimeter wave (mmWave), non-orthogonal multiple access (NOMA), massive multiple-input and multiple-output (MIMO), terahertz (THz), and visible light communication (VLC); MEC provides the Metaverse with massive server resources to power the virtual 3D realm, while providing a network of ultra-low latency communications for imperceptible response times; Digital twins can create a virtual replica of physical objects or services, which provides users an immersive experience in the physical world.}
\label{fig_4pillars}
\end{figure*}

\subsection{Responsible AI}
\subsubsection{Motivation}
While AR/VR is at the forefront of the Metaverse, AI is an important technology that works behind the scenes to make miracles happen. AI is most useful for data computations and predictions. It can help improve algorithms for tasks such as avatar creation, natural language processing and translation, and world generation. It could also improve how we interact in VR, as AI monitors sensors that measure our bioelectricity and muscle patterns. AI also makes experiences more inclusive by offering services such as image recognition for visually impaired users.

As many AI techniques and applications bloom in the Metaverse, it becomes more and more critical to comprehend and retrace the functioning of the applied AI techniques. This can ensure that the Metaverse is working as expected to meet regulatory standards. 
Due to the high complexity of the Metaverse's construction, the learning process is turned into what is commonly referred to as a ``black box'', which is difficult to interpret. For example, deep neural networks are some of the hardest for a human to understand. These black box models are created directly from the data. 
As a result, data processing in the Metaverse can suffer data biases, resulting from population data, measurement error, data quality chasm, data re-purposing, and data augmentation. Such data biases can jeopardize data quality, fairness, accountability, transparency, and explainability of the Metaverse. Despite a proactive policy that can be used to build trust and transparent AI for fighting against negative social impacts, developing such a policy craves a deep understanding of AI techniques in the Metaverse, where the AI decision-making processes have to be explainable~\cite{madhavan2020toward}. 

\begin{figure}[htb]
\begin{center}
\includegraphics[width=0.5\textwidth]{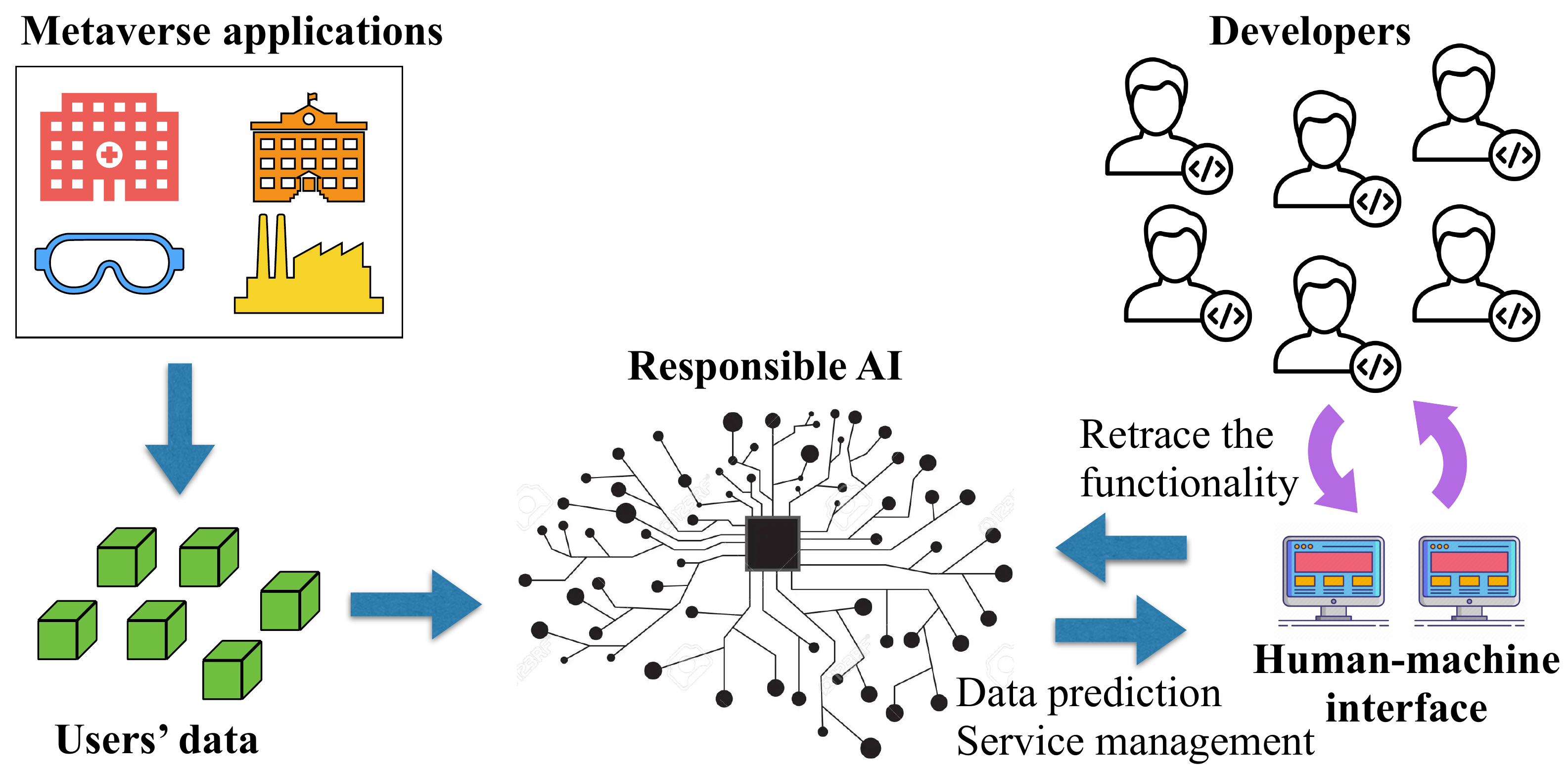}
\end{center}
\caption{Responsible AI can enable a full understanding of the AI decision-making processes in the Metaverse.}
\label{fig_resAI}
\end{figure}

\subsubsection{Significance}
To promote the trust and usability of the Metaverse, responsible AI~\cite{arrieta2020explainable}, as shown in Figure~\ref{fig_resAI}, is developed to fully understand the AI decision-making processes while considering the ethical, moral, legal, cultural, and socio-economic consequences in the Metaverse. Moreover, the responsible AI helps the engineers or data scientists who develop AI algorithms explain what is exactly happening inside the IoT applications and how the AI algorithm leads to a specific result~\cite{werder2022establishing}. Responsible AI can be applied to various applications, which extends the responsible guidance on the limits of a learning algorithm, and describes the training data involved~\cite{wearn2019responsible}. In~\cite{yigitcanlar2021responsible}, a responsible urbanization framework is studied with the AI system, which aims to balance the benefits, risks, and impacts of developing the city. Responsible AI can consider the effects of machine learning decisions on the natural environment and future urban planning. Accordingly, responsible urban innovation can be characterized by sustainability, which is important for planning long-term urbanization strategies. 

\subsubsection{Key technologies}
Several explainable AI-based approaches~\cite{mohseni2021multidisciplinary} are introduced in recent years to strengthen control and oversight in case of adverse or unwanted effects, such as biased decision-making or social discrimination. 
In~\cite{thakker2020explainable}, deep learning is used to classify drainage and gully images for flood monitoring in smart cities, where the image can be labeled to one of the three blockage tags (No Blockage, Partially Blocked, and Fully Blocked). Explainable AI combines semantic representation to unveil relationships between various objects' coverage within the image, thus improving the explainability of deep learning solutions. To assist the product design and marketing, explainable AI can be used to characterize competitors' work and identify potential opportunities to outperform them~\cite{han2021explainable}. Based on customer-generated reviews, explainable AI can extract the competitive factors of the product and effectively reflect customer opinions. 

\subsubsection{Applications in Metaverse}
Responsible AI has been widely used for building smart health systems in the Metaverse. The capabilities of responsible AI can drive the healthcare market and enable patients to experience a dynamically different service environment~\cite{kumar2021responsible}. Data sufficiency and integrity, algorithmic design, privacy protection, malfunction correction, and collaborative efforts are critical factors in shaping responsible AI with the smart health systems in the Metaverse. 
The authors in~\cite{vourganas2020individualised} develop responsible AI with a patient-centric individualized home-based rehabilitation support system. Synthetic datasets complement experimental observations and mitigate bias, where explainable AI-based algorithms combine ensemble learning and hybrid stacking using extreme gradient boosted decision trees and k-nearest neighbors to meet individualization, interpretability, and system design requirements. 
Socially responsible AI~\cite{cheng2021socially} is studied to prioritize the data and the users of the Metaverse while eliminating the negative affection of minoritized or disadvantaged users. Based on the users' feedback, socially responsible AI accomplishes the expected social values of the data to maximize the long-term beneficial impact. 
In~\cite{khodabandehloo2021healthxai}, explainable AI is extended to recognize early symptoms of cognitive decline, which explains why a specific prediction is generated. The explainable AI-based system is built with clinical indicators describing subtle and overt behavioral anomalies, spatial disorientation, and wandering behaviors. The explainable AI-based system is adaptive to different individuals and situations to recognize anomalies. 

Enabling the  Metaverse in industry 4.0 often requires hyperparameters optimization, finetuning, computing capability, and continuous training of huge data. The data are typically generated and aggregated from millions of devices. Thus, it is possible that the users' sensitive data can be falsified due to cyberattacks. Thus, more and more applications in industry 4.0 explore explainable AI with Metaverse to strengthen the risk assessment, trustworthy control operation, and automation~\cite{ahmed2022artificial}. 
Furthermore, a risk assessment framework~\cite{al2021explainable} is developed for human-sensing IoT systems, where human attention states are classified and mapped into four categories, i.e., low, normal, medium, and high risks. Explainable AI is developed to classify abnormal electroencephalogram (EEG) brainwave signals based on the importance of each feature used in the deep learning model. In IoT-based smart agriculture, IoT nodes are deployed in remote farms and sense metrics, such as temperature or humidity. The sensory data is used to adapt irrigation control systems. For precise crop monitoring and automated irrigation, explainable AI can be utilized to produce interpretable data analysis~\cite{tsakiridis2020versatile}, which enables users to inspect the model reasoning in the irrigation control systems. 

Taking advantage of 5G networks, COVID-19 can be detected based on uploaded patients' medical images or blood samples. In~\cite{hossain2020explainable}, deep learning models are trained at the 5G servers to process the medical data, where explainable AI is used to allow doctors to understand the outputs of each layer of the deep learning models. In the explainable AI, multi-modal datasets, such as CT/X-rays images and ultrasounds, as well as protease sequence data, confidence scores, and configuration settings of the deep learning model, are fused to generate a knowledge graph. The knowledge graph is able to reveal the data disruption and the feature adjustment probability. The explanations can also be visualized by a data visualization method, e.g., the gradient-weighted class activation mapping (gradCAM). 
Explainable AI is promising to quantify the uncertainty of 5G and beyond networks, especially understanding training data and AI algorithms concerning the wireless network performance requirement~\cite{guo2020explainable}. Particularly, the data feature visualization of the explainable AI can improve the transparency of AI-enabled network layers, which ensures mission-critical services in 6G networks. 
Explainable AI also motivates interactions between AI-based human-centric systems and 6G networks~\cite{wang2021explainable}, where explainable AI aims to build trust between end users and AI-based machines in radio and resource allocation as well as network optimization.

\subsection{High-speed data communications}
\subsubsection{Motivation}
The Metaverse aims to provide massive mobile users the connected AR/VR experiences and real-time services. Providing a networking AR/VR service relies on enormous amounts of data exchange with ultra-low latency, where the nature of haptic signals and human perception requires 1 ms of latency at maximum. However, most 4G technologies cannot supply real-time data communications for multiple mobile users. For example, it is hard to reduce the latency of LTE-based wireless systems below 25 ms~\cite{sukhmani2018edge}. 

In recent years, 5G is rapidly developed to provide real-time data exchange to IoT devices~\cite{zhao2020edge}. The data rate of 5G outperforms 4G-based systems while providing a suitable high-speed communication architecture for the Metaverse. In particular, 5G networks enable several promising technologies, such as mmWave~\cite{lin2020tensor,mir2022relay}, NOMA~\cite{xiao2019downlink,zeng2019ensuring,wang2021joint}, and massive MIMO~\cite{krogfoss2020quantifying}. Based on the large-scale commercial deployment of 5G networks, 6G is expected to extend personalized communications to fully realize the machine-to-machine paradigm, which connects not just people, but also IoT devices, smart vehicles, wearable sensors, and even mobile robots~\cite{giordani2020toward,chen2021code}. These unique additions within high-speed data communications can empower millions of connected devices and ubiquitous AR/VR applications, which allows the Metaverse to benefit from the ``anytime-anywhere connectivity'' promise of next-generation wireless mobile networks. 

\subsubsection{Significance}
High-speed data communications can improve the reliability of the services as well as the network capacity and density of the Metaverse. Specifically, 5G can boost the data rate of the IoT device or the mobile user to 100 Mbps in a uniform spatial distribution, while the peak data rate can be 10--20 Gbps~\cite{liu2020toward}. The high data rates support enhanced mobile broadband for personal mobile services and large-scale machine-to-machine communications. Moreover, 5G offers reliability and delay critical services, where the end-to-end latency is maintained as low as one millisecond, and the reliability is as high as 99.99\%. Compared with the current 4G techniques, 5G networks provide a 10-time increase in the network throughput, a 10x decrease in communication latency, a 100x enhancement in the traffic capacity, and a 100x growth in the network efficiency. 

Based on the existing 5G architecture, 6G can be developed to fulfill the ubiquitous application requirements in the Metaverse. For example, providing the enhancement of VR/AR capabilities with the required resources (such as computing power, storage space, graphics processing capabilities, and communication resources) through massive IoT devices~\cite{liao2020information}. For ultra-high data rate and further reducing communication latency, 6G is studied to utilize more frequency bands than 5G and 4G, such as sub-THz and THz, as well as VLC. Although it is well-known that a high radio frequency can result in an increased data loss, 6G benefits from the shortened data transmission distance to support more low-latency applications in the densely-populated area. To address the issue of limited coverage, 6G also applies high-speed and low-latency machine-to-machine communication techniques and ultra-massive MIMO~\cite{zhang2020envisioning}.

\subsubsection{Key technologies}
To support several gigabits-per-second high-speed data rates, the working frequency of the mmWave IoT systems is set between 30 and 300 GHz~\cite{sun2018propagation,prabhakara2020osprey}. Some of the works studied use large-scale antenna arrays for the high-frequency band of the mmWave~\cite{zhang20165g}. The antenna elements can be arranged linearly or in a full-dimensional array (i.e., with both elevation and azimuth angle resolution capabilities). Due to the high carrier frequency, the mmWave IoT systems suffer a more severe propagation loss than the low-frequency communication technologies, such as LTE. Moreover, it is difficult for the mmWave signal to penetrate and diffract around obstacles, such as vehicles and buildings in urban areas, and molecular (e.g., rain and dust) and atmospheric (e.g., air density) absorption can also prevent the mmWave propagation. 

Another 5G technology is NOMA which shares non-orthogonal time and frequency resources via power-domain or code-domain multiplexing. The power-domain multiplexing allocates the IoT devices with different transmit power adapting to the time-varying channel quality to achieve a maximized network benefit. Moreover, IoT devices can conduct code-domain multiplexing, which generates different codes that can be multiplexed over the same time-frequency resources. Although the code domain multiplexing may consume more signal bandwidths than power-domain multiplexing, the former can improve spreading gain and shaping gain. NOMA can sustain massive connectivity in the Metaverse, where the number of orthogonal resources available does not restrain the number of IoT devices. 

\begin{figure}[htb]
\begin{center}
\includegraphics[width=0.5\textwidth]{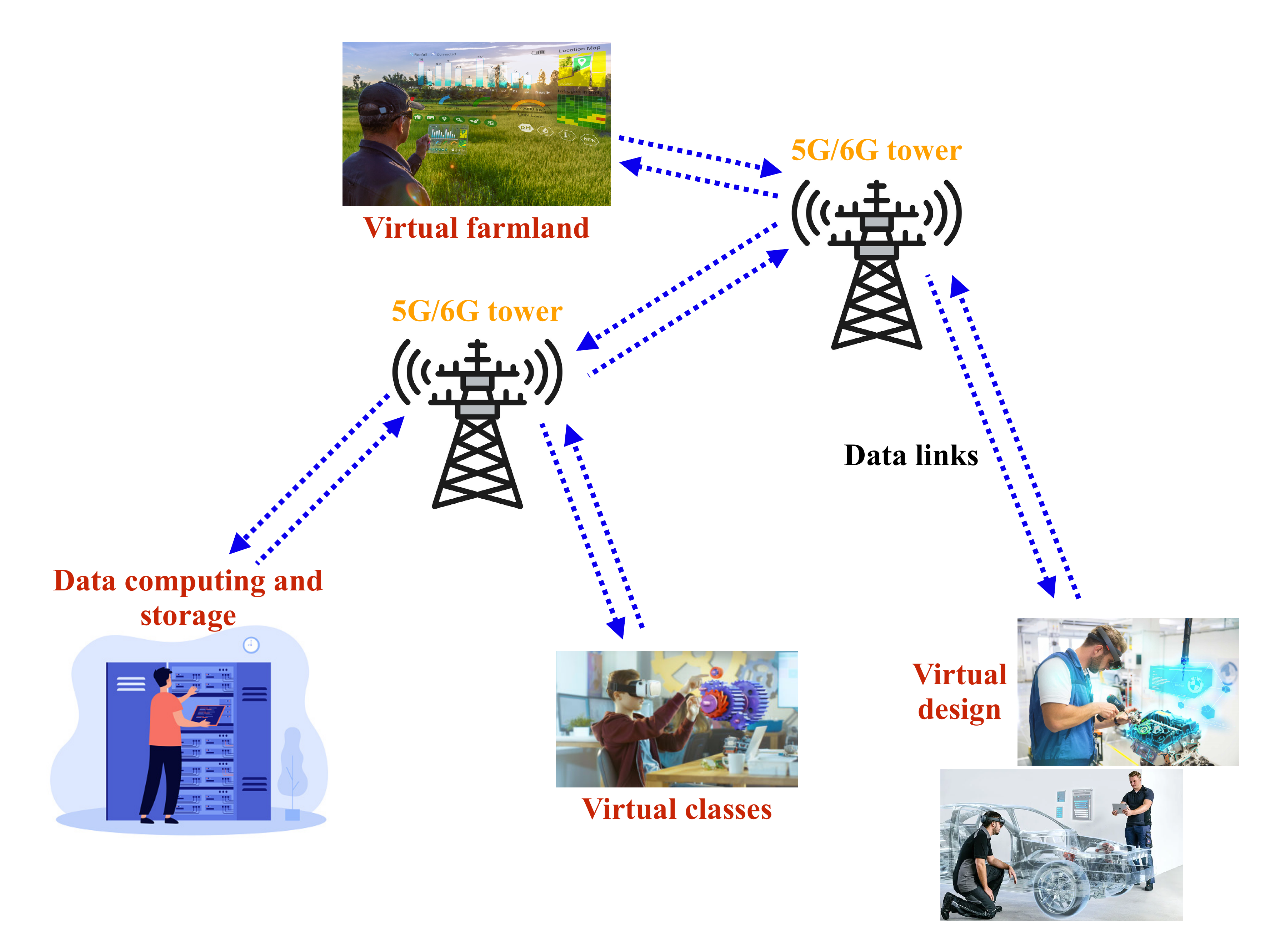}
\end{center}
\caption{The high-speed communication technology enables ``anytime-anywhere connectivity'' for the Metaverse applications, where the AR/VR can be used for virtual design and manufacture, teaching, and farmland monitoring.}
\label{fig_5g6g}
\end{figure}

To meet the demand of 100 Gbps data rates and 1-millisecond latency, THz of 6G can be applied to IoT networks that facilitate massive AR/VR service coverage and scalability in the Metaverse. Specifically, the range of THz is from 300 GHz to 3 THz since most of the spectrum beyond 100 GHz is still idle. As new technology develops, THz communication is still facing obstructions. For example, THz signals can be easily attenuated by weather influences, such as raining, snowing, or even water vapor in the atmosphere. Additionally, the propagation loss increases quadratically with the frequency of long-distance communication. This can result in a severe free space path loss in the THz communication, e.g., a 10-meter THz link suffers 100 dB loss. 

\subsubsection{Applications in Metaverse}
Figure~\ref{fig_5g6g} presents the high-speed data communications in the Metaverse, which provides ``anytime-anywhere connectivity'' for the IoT applications, such as virtual design and manufacture, teaching, and farmland monitoring. Specifically, high-speed data communication allows students to remotely attend virtual classes, where teachers anywhere on the planet vividly demonstrate the content knowledge via the video/audio stream. The virtual classes can moderate student participation, enrich the learning experience, and break down the most common barriers that hinder synchronous learning. 

The high-speed data communication can also enable engineers in different industrial manufacturing companies to jointly design their products online. Engineers can utilize IoT data and AR/VR to create a shared virtual 3D model. The Metaverse that takes advantage of high-speed data communication brings complementary technologies and industrial design patterns to integrate the physical and cyber worlds. Moreover, the industrial company focusing on hardware design can connect with hundreds of certified partners who provide software support or supply chain solutions. This will eventually lead to a comprehensive platform for virtual 3D design and collaboration with physics-based digital models. 

\begin{figure}[htb]
\begin{center}
\includegraphics[width=0.5\textwidth]{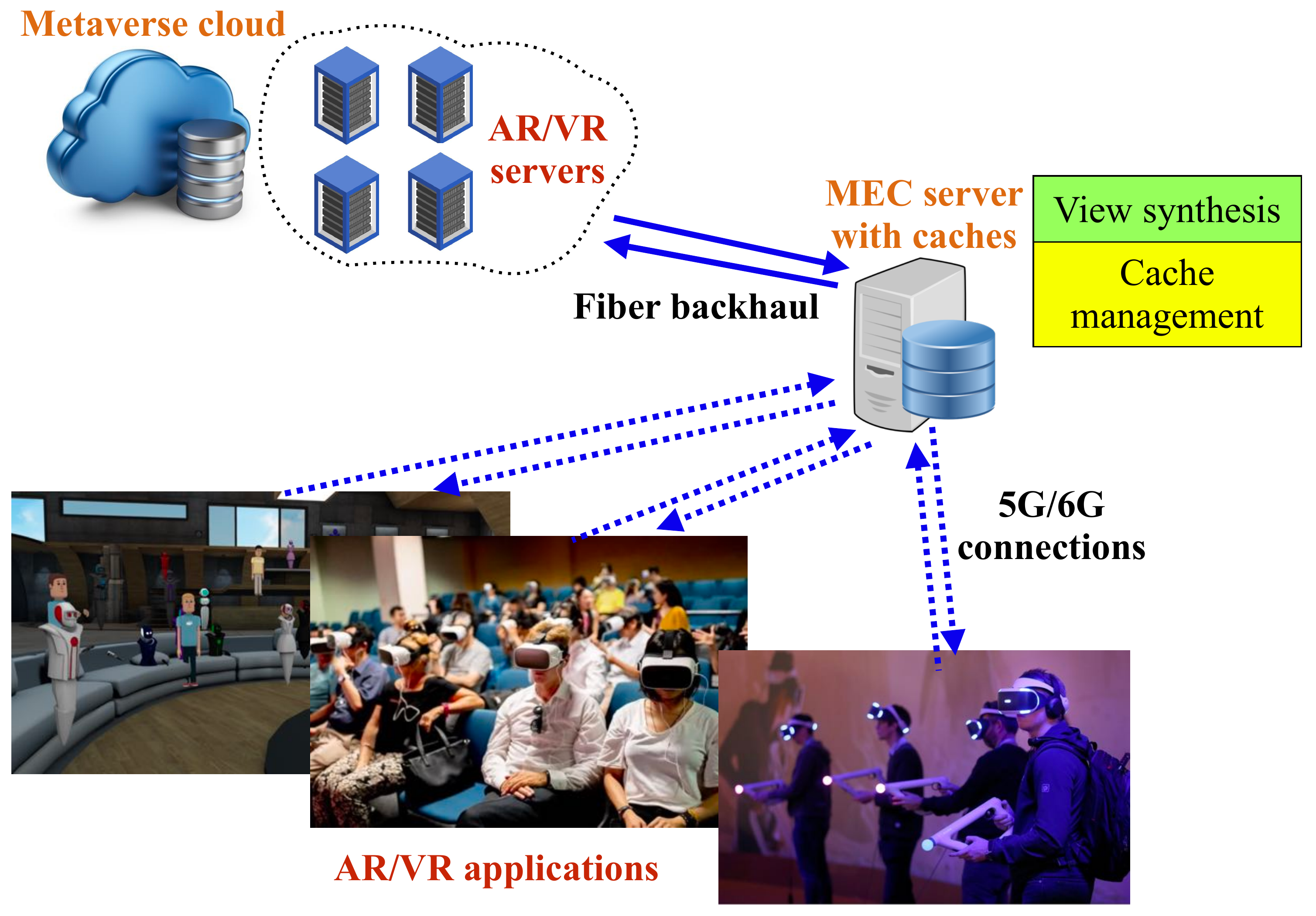}
\end{center}
\caption{By caching the video contents at the edge server, MEC offers rapid response time to the computing resource request from the users' Metaverse application. Particularly, the requested data of the user can be fetched directly from the edge cache/the cloud cache/the AR/VR video source server or synthesized by the MEC server~\cite{dai2019view}.}
\label{fig_MEC}
\end{figure}

\subsection{Cost-effective Mobile Edge Computing}
\subsubsection{Motivation}
It is critical to maintain user engagement and immersion in the Metaverse to provide the same level of experience as in reality. In this regard, the response time to user actions should be significantly reduced to levels below the human perceptible limit.

Mobile Edge Computing (MEC) is an emerging technology that provides fast response times by hosting a combination of communications infrastructure and computing resources close to the users. As shown in Figure~\ref{fig_MEC}, MEC extends a network and software platform environment that can seamlessly approach or even exceed the limits of human perception of response time.
The dynamic and immersive experience requirements of the Metaverse can be achieved by adopting MEC. MEC can provide the Metaverse with massive server resources to power the virtual 3D realm while providing a network of ultra-low latency communications for imperceptible response times.

\subsubsection{Significance}
As 5G (expected to reduce last-mile latency to 1ms) and future 6G flourish, MEC is a promising technology for the Metaverse. It can provide single-hop edge offloading services from cellular-connected user devices to enhance user experience in the  Metaverse, such as AR glasses. MEC is also essential for outdoor Metaverse services to understand the detailed local environment and coordinate close collaboration between nearby users or devices. For example, 5G MEC servers can manage AR content for nearby users via single-hop packet transmission and enable real-time user interactions for social AR applications, e.g., ``Pokémon GO''~\cite{chen2021ar}.

\subsubsection{Key technologies}
MEC is implemented based on the following three key components:
\begin{itemize}
    \item Edge offloading and Computing: In addition to offloading rendering computations to the edge nodes, expensive computation tasks (such as matrix multiplication) required for data processing and AI model training, can also be broken down into subtasks and offloaded to edge servers. The computation results can be resumed by aggregating all the completed subtasks at a central node.
    
    \item Edge Caching: Edge caching helps reduce computational and communication redundancy, by reducing repeated access to popular content or computations.
    In the Metaverse, the probabilistic models of file popularity distributions (such as field of view) can be learned. The popular field of view can then be stored on edge nodes close to users who need more reduction in computational cost and rendering latency.
    
    \item Privacy-preserving Local model Training: As more users connect to the Metaverse, the risk of data breaches increases as the attack surface increases. Federated Learning (FL) is one of the solutions to protect user privacy~\cite{yang2019federated}. 
    FL represents an approach of distributed machine learning and learns a global model by aggregating users’ uploaded models trained on local devices without directly leaking private information.
\end{itemize}

\subsubsection{Applications in Metaverse}
Researchers have proposed several solutions to improve the performance of Metaverse applications by using the latency advantages of edge computing.
For example, the MEC was introduced into the Metaverse to improve the quality of user experience~\cite{zhang2017towards},
where dynamic edge nodes in the MEC architecture are as close to users, and multiple edge nodes can be combined to assist in completing the same user's instructions, effectively solving the delay issues caused by user mobility in the Metaverse.
However, the cooperation of multiple edge nodes requires the transmission of user data, which may lead to serious consequences such as user privacy leakage and identity crisis in the Metaverse.
An FL-based MEC architecture for the Metaverse was proposed in \cite{tuli2021mcds} to protect users’ privacy, where locally trained models are transmitted between edge nodes instead of user data.
Furthermore, a directed acyclic graph was established by the edge nodes to improve the system's computational efficiency.

The use of MEC to enhance the Metaverse experience has also attracted wide attention from the academic community.
Dai et al. \cite{dai2019view} designed a 360-degree VR caching system based on view synthesis over MECCache servers in Cloud Radio Access Network (C-RAN) to promote the QoE of wireless VR applications. Both Gu et al. \cite{gu2021reliability}, and Liu et al. \cite{liu2018mec} exploited sub-6 GHz and mmWave links combined with MEC resources to deal with the constrained VR HMDs resources and transmission rate bottlenecks of normal VR and panoramic VR video (PVRV) transmission.

\begin{figure}[h]
    \centering
    \includegraphics[scale=0.3]{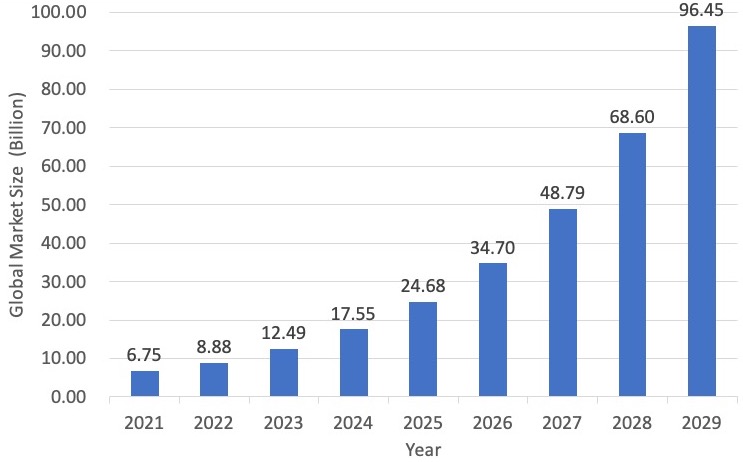} 
    \caption{Prediction of digital twins market size.}
    \label{fig:dt_markets_size}
\end{figure}

\subsection{IoT-based digital twins}
\subsubsection{Motivation}
Digital twins create a virtual replica of physical objects or services, such as building facilities, operational processes, human-computer interaction, and social services. In the Metaverse, digital twins are typically used to provide an immersive shopping experience in the physical world. Combining digital replicas with physical products and services can also support data analytics, allowing commercial individuals to utilize physical-world scenarios for simulations before making costly decisions. As digital representations of physical-world entities, digital twins can synchronize assets, processes, and operating systems with the real world and everyday actions such as visualization, analysis, and prediction~\cite{Tao2019digitaltwin}. The digital twins are central to how the physical and virtual worlds interact through the IoT connections~\cite{Chen2021digitaltwin}. Therefore, changes in the physical world are refused in the digital world. These unique digital twins could be one of the fundamental building blocks of the Metaverse, creating replicas of the physical world, including its structure and functionality, to serve as gateways for users to access and enjoy virtual services. For example, an engineer could create a 3D representation of a complex system at different levels of complexity (i.e. descriptive, informational, predictive, holistic, autonomous) for different functionality. Digital twins enable the engineers and service providers to clone virtual objects of machines and processes and perform physical analysis  remotely via AI~\cite{Rathore2021digitaltwin}.

Since the concept of digital twins was first proposed by David Gelernter in his book\cite{noauthor_mirror_1993}, digital twins have given businesses an unprecedented view of how the products are designed, operated, and performed, which helps make it easier to provide better products or service.

According to Fortune Business Insights~\cite{FBI}, the global digital twins market size has reached  \$6.75 billion in 2021 and is projected to reach \$96 billion in 2029, at a compound annual growth rate (CAGR) of 40.6\% during the forecast period, as shown in Figure~\ref{fig:dt_markets_size}, a major driver is NVIDIA Corporation unveiling a new platform to expand the user base, while COVID-19 restricts industry development. 

\begin{figure}[h]
    \centering
    \includegraphics[scale=0.47]{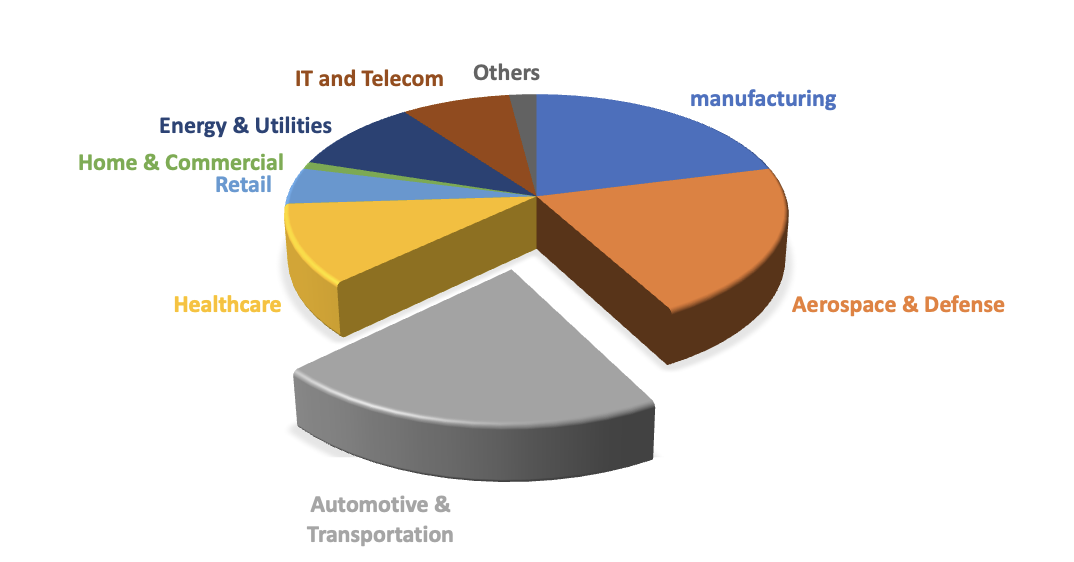}
    \caption{The global market share of the digital twins in 2021.}
    \label{fig:dt_markets_share}
\end{figure}

The global market share based on end user in 2021\cite{FBI} is illustrated in Figure~\ref{fig:dt_markets_share}. Among end users, the digital twins applications that occupy the largest share are automotive and transportation. The reason is mainly due to breakthroughs in advanced technologies such as AI and the continuous development of the commercial value of autonomous vehicles. In addition, industrial manufacturing and aircraft manufacturing account for a high proportion, both exceeding 20\%. The market share of digital twins in IT and communication, healthcare, and smart homes is around 10\%, which are all emerging digital twins applications and have great development prospects.

Many businesses, such as NVIDIA\cite{NVIDIA}, are investigating the correlation of the Metaverse and digital twins, introducing new possibilities and experiences for digitally-driven consumers. \textit{LeewayHertz}~\cite{leewayhertz} believes that as more companies and enterprises use digital twins to build products and digital Apps in the Metaverse, they can create entire ecosystems over time. Due to the high development cost, the existing digital twins are mainly used for the construction of complex, multi-level large-scale industrial products or systems, such as manufacture~\cite{manufacture}, healthcare~\cite{9320532}, engineering, and smart cities~\cite{deren_smart_2021, LI2022167}. 

\subsubsection{Significance}
Digital twins produce a digital representation of a physical-world product or service. Implementing digital twins is an encapsulated software object or model, where multiple IoT data are aggregated to create composite views across many physical-world entities. When we take an object in the physical world and create an exact copy of the same object in the digital world, this system allows users and engineers to mimic the behavior of real objects and predict their movements and reactions. Digital twins, therefore, enable strategic product development while avoiding the potential for costly mistakes. Digital twins are one of the core building blocks of the Metaverse due to their unique nature. The IoT-empowered Metaverse helps people create virtual worlds and experiences beyond the wildest dreams,  and build exact replicas of reality that can bring reality to the digital world.

Digital twins are of little value unless IoT networks are updated with real-time data, a task that machine learning and AI have grappled with for quite some time. The Metaverse is where IoT and digital twins gather and celebrate, and users look forward to using some cool AR/VR applications and experiencing immersive services in 2022 and beyond. First, we have to correct a misconception, namely, the idea that the virtual scan object is the digital twins of the Metaverse. A scanned object can be the outlook of digital twins. Still, it can be used in digital twins only when the actual metadata \& physical-world data feed from IoT devices are overlaid simultaneously. Linking physics objects in digital form with overlay data is significant in complex simulations.

\begin{figure}[h]
    \centering
    \includegraphics[scale=0.4]{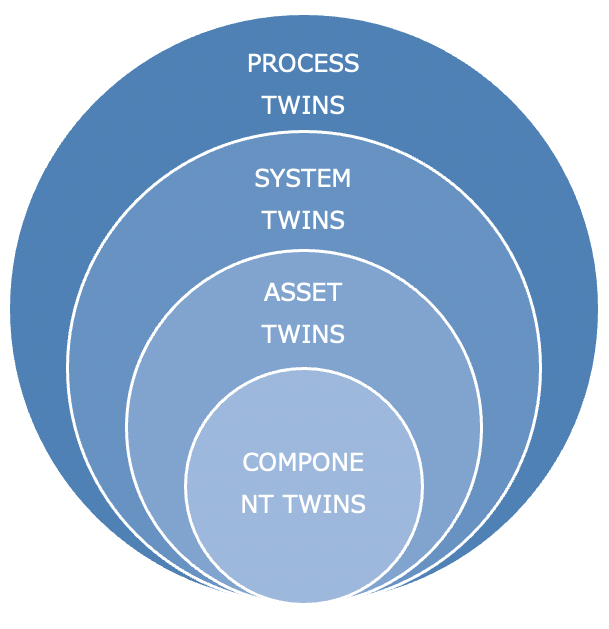}
    \caption{The hierarchical architecture of digital twins.}
    \label{fig:dt1}
\end{figure}

\subsubsection{Key technologies}
Digital twins can be divided based on their scale and inclusiveness into component twins, asset twins, system twins, and process twins~\cite{Fuller2020digitaltwin}, as illustrated in Figure \ref{fig:dt1}. We also summarize the correspondence between different applications and this development architecture in Table \ref{table_applications}.

\begin{itemize}
    \item[1.] Component twin: The smallest part of the entire system is the component twins~\cite{kapteyn_data-driven_2022}, which are composed of key components that affect performance, or other components with unstable performance To avoid data redundancy and reduce costs, unimportant components do not need to be replicated. Azure Digital Twins (ADT)~\cite{ADT} is a platform developed by Microsoft, which not only allows the creation of models but provides a graph API to query and interact with its twin. Basic examples of a series of 3D component models of the devices, machinery, infrastructure, and buildings already exist.
    \item[2.] Asset twin: Asset twins are higher level than component twins~\cite{kapteyn_data-driven_2022}. An asset twin is a collection of information from a component twin, or the component twin itself. Asset twins identify the potential for improvement and turn them into actionable insights by analyzing performance data, which results from the interaction of these components. GE Healthcare~\cite{GEhealthcare} states that some asset twins have been used to solve healthcare problems in hospitals, such as staffing model design and surgical block schedule optimization.
    \item[3.] System twin: System twins operate at the next level, where different assets are combined to form a fully functional system, such as a brake system in automobile~\cite{manufacture}. System twins provide visibility into asset interactions, further enhancing performance~\cite{RUDSKOY2021927}.
    \item[4.] Process twin: Process Twin leverages high-performance computing methods to optimize equipment and the entire manufacturing process by integrating multi-dimensional process knowledge models~\cite{noauthor_digital_nodate}. By integrating production procedures and economics, manufacturers can achieve unprecedented efficiency and insight.
\end{itemize}

\begin{table*}[h]\centering
\caption{Hierarchical evolution}
\begin{tabular}{c|p{2.8cm}p{2.3cm}p{2.3cm}p{2.5cm}}
\hline
\diagbox{Scenarios} {Architecture}   & Component twin                                                                 & Asset twin       & System twin      & Process twin           \\ \hline
Manufacture & Sensor, screw  ~\cite{manufacture}                                                                      & Engine, pump~\cite{manufacture}  & Electricity, brake~\cite{manufacture} & Collaboration ~\cite{manufacture}          \\ \hline
Supply chain     & Shipments    ~\cite{shoji_optimizing_2022}                                     &Supply\newline  environment ~\cite{WANG2020198}                             & Individual\newline  stores    & Chain \newline improvement   ~\cite{shoji_optimizing_2022}   \\ \hline
Healthcare & Surgical equipment          & Hospital~\cite{austin_architecting_2020}          & Medical \newline system~\cite{croatti_integration_2020}    & Emergency rescue   ~\cite{9320532}      \\ \hline
Smart city& Infrastructure~\cite{mohammadi_knowledge_2020} &  Campus~\cite{mohammadi_knowledge_2020},\newline  hospital~\cite{austin_architecting_2020}& Small\newline  community   & Efficient city\newline  operation~\cite{deren_smart_2021, LI2022167} \\ \hline
\end{tabular}
\label{table_applications}
\end{table*}

\subsubsection{Applications in Metaverse}
{ To monitor and analyze welder behavior, a digital twins system has been designed with VR and AI technologies in~\cite{Wang2021digitaltwin}.
In \cite{Elayan2021digitaltwin}, a data-driven digital twins framework was investigated to improve health diagnosis performance and operation in smart medical systems.
Digital twins were designed in \cite{Ghandar2021digitaltwin} to replicate a virtual representation of agricultural production, where ML algorithms process the sensory data obtained from actual sensors in decision support systems.

In California, a visualization of wildfire areas was created. 
Fire behavior analysts can use visualizations and data for predictions that allow firefighters to better extinguish real-world fires by taking data from simulated experiences without injury.
To detect production errors and mistakes, enterprises create digital twins of their factories and manufacturing plants. 
Visualizing all aspects of production helps optimize the design, manufacturing, and labor processes.
Retail and e-commerce companies allow customers to shop at virtual storefronts at home. 
Product availability is updated in real-time, and consumers could interact with the items they intend to buy.
In the Metaverse, client meetings could be as simple as a client visiting your virtual headquarters and sitting in your virtual conference room, eliminating the need for travel, while providing people with a face-to-face experience.  
Enriching digital objects with connections to data sources within the Metaverse can provide teams with every piece of information they need about the companies or individuals they will encounter to gain an edge in meetings and proactively help them.
}
 
Metaverse uses digital twins throughout the product development life cycle, from designing to post-production monitoring and service, to improve productivity and profitability~\cite{SCHLEICH2017141,boschert_digital_2016}.
During the product ideation phase of Metaverse, digital twins can be used to build virtual prototypes for Metaverse~\cite{malik_digital_2018} and generate large amounts of predictive data about performance outcomes. Before investing in a physical prototype, make necessary product improvements based on forecast results to meet pre-set requirements. This helps businesses save time and cost by reducing the number of iterations required to bring a Metaverse product into production.

During the development of Metaverse products, digital twins help improve the development process in the following ways, including process optimization, remote device diagnostics, and supply management. For example, digital twins help identify potential failures, troubleshoot remotely, and ultimately improve customer satisfaction~\cite{kenett_digital_2022}. In supply chain optimization, the production timing and supply routes can be optimized by digital twins~\cite{shoji_optimizing_2022}. 
New methods can be discovered by adjusting the digital twins to optimize production, reduce variance, and aid in root cause analysis.

Digital twins can also be used in later stages for metaverse products, such as cost reduction, predictive maintenance, and service improvement. When commercializing  metaverse products, digital twins can collect user data over time to gain insights into product performance and service experience~\cite{jiang_industrial_2021}. This data can help engineers and designers improve the customer experience through customized services.

%=============================================================================%
%===================================Section 5 Challenges==========================%
\section{Open Issues in the Future IoT-empowered Metaverse}
\label{sec_cha}
Despite the studied four pillar technologies can be leveraged to empower the Metaverse with IoT, further development and implementation of IoT-empowered Metaverse need to address the following critical open issues.

\subsection{Data processing}
The Metaverse applications, such as AR and VR, need to collect real-time IoT data to analyze the behaviors or states of the physical world objects or procedures~\cite{ghosh_developing_2021}. How to monitor equipment and collect data from all aspects, including initial data, monitoring information, operational information, business data, and program data, are very challenging~\cite{8477101}. 

Once the real-time IoT data is collected, the data mining and data analysis have to be performed. Unlike traditional simulations~\cite{kenett_digital_2022}, the Metaverse can run multiple simulation processes with real-time data and give real-time feedback to the source object for improvement. These simulation processes rely on data modeling techniques, such as engineering simulations~\cite{boschert_digital_2016}, physical analysis~\cite{jiang_industrial_2021}, machine learning~\cite{austin_architecting_2020}, neuromorphic computing, and data mining~\cite{jiang_industrial_2021} to help users/engineers have a full view into how the virtual world is performing. 

Spatial computing~\cite{shekhar2015spatial,greenwold2003spatial} can be applied for the human interaction with the IoT devices, where the IoT devices retain and manipulate referents to real objects and spaces. To utilize the spatial computing, the Metaverse needs to ensure the data scalability and diversity, for example, room temperature, service locations, AR/VR equipment details, users' health conditions, social connections, hobbies and interests. Moreover, most commercial AR/VR equipments have various sensors that generate a huge amount of data with different modalities and types. 
Note that the Metaverse can take advantage of data fusion techniques from different data sources to regenerate a comprehensive view (e.g., car driving and road condition), rather than multiple separate perspectives~\cite{8477101,liu_role_2018}. Data management issues such as storage, transmission, security, and privacy of dense information are critical. Constructing a standardized semantic representation of different data~\cite{9380190}, which is used to communicate between intelligent machines inside the Metaverse, is a good way to compress transmission information, reduce transmission delay and encrypt the original data. In addition, ensemble learning techniques can also be considered for data fusion. Data fusion is also a key to the Metaverse to enhance the interoperability between different platforms and systems~\cite{lim_realizing_2022}.

\begin{table*}[h]\centering
\caption{Data processing}\label{dataprocessing}
\begin{tabular}{l|p{5cm}|p{6cm}}
\hline
\textbf{Requirements} & \textbf{Challenges}                               & \textbf{Potential Technologies}                                                                      \\ \hline
Data Collection      & Real-time sensory data, Data transmission                 & IoT network, Sensor network                                                                         \\ \hline
Data Modeling        & Data mining, Data analysis, and Real-time feedback & Engineering simulations, Physical analysis, Machine learning, Neuromorphic computing, and Data mining \\ \hline
Data Fusion          & Storage, Transmission, Security and privacy       & Intelligent Machine Communication Network, Semantic Communication, Ensemble Learning                                    \\ \hline
\end{tabular}
\end{table*}

\subsection{Security and privacy}
In the Metaverse, shared user data and 3D virtual spaces are interweaved with the IoT system and merged for remote virtualization. The Metaverse  is very likely to integrate services, applications, or IoT systems from multiple companies or institutions~\cite{kuppa2020black}. As a result, the major security and privacy concern are due to the fact that the integration of different IoT devices requires all partners to coordinate and interact data with each other. 

The responsible AI is expected to help the Metaverse developers and designers understand the ``black box'' of AI, which improves the integration efficiency of the IoT systems. However, these explanations also expose extra information about the AI decision-making process that utilizes sensitive users' private information. Such a privacy leakage can be utilized by the attacker, for example, to identify certain users and their health conditions given attributes such as gender, race, and birth date. The attacker can also inject a relatively small amount of poisoned data into the AI model. Since the AI model is typically used to process the ``big data'', detecting the poisoned malicious data in the Metaverse application is difficult. With the growth of the training time, such inconspicuous poisoned data can affect more and more functions in the AI model, eventually resulting in Metaverse service failures. 

Furthermore, the human-machine interface poses security and privacy risks to the responsible AI. Especially when the Metaverse is applied to autonomous vehicles and industrial manufacturing. Due to the dynamic application environment, it is difficult for the Metaverse developers and designers to recognize when the AI model should be overruled, or the human's decision can override the AI model. Human judgment can prove faulty in overriding system results, such as scripting errors, lapses in data management, or misjudgments in model-training data. Therefore, rigorous safeguards at the human-machine interface are critical for the responsible AI, which can protect the Metaverse from corrupting algorithms or using the AI model in malfeasant ways.

The Metaverse, as it is imagined today, requires the transfer of huge amounts of personal and/or critical data~\cite{cai2022compute}. Security, privacy, and trustworthiness will be more difficult in the Metaverse than in the physical world~\cite{wang2022survey}, thus providing the technical and social means to assert rights and enforce regulations may not get any easier. The Metaverse relies on IoT data to operate. Cyber-attacks and data storage are the major challenges for deploying the Metaverse applications like gaming, socializing, real estate, and healthcare~\cite{menhaji2022systematic,bian2021demystifying}. The cybersecurity of the Metaverse interfaces has to be secured to create a sustainable Metaverse that can be used in the long term.

\subsection{Real-time 3D modeling}
One of the critical components in the Metaverse is 3D modeling, which aims to stream the AR/VR experiences like streaming a movie. Typically, AR/VR-based computer games require powerful gaming computers to process the data as fast as possible. Moreover, it is a bottleneck in the communication networks to provide fast access to many AR/VR online games in parallel. The data exchange delay between the IoT devices (generally measured in milliseconds (ms)) determines the number of frames per second (FPS) streaming to an IoT device. If the IoT devices slow the frame rates to 15-30 FPS, when people turn their head up and down, or left and right, it might make the user feel dizzy in the AR/VR experience. Therefore, 15-30 FPS is the minimum for most AR/VR games. In addition, the frame rates normally should be set to 60 FPS, which is ideal for AR/VR games. 90 FPS is usually a comfortable standard and a bare minimum for VR. In particular, current 4G technology supports around 50ms (about 20 FPS) on average for data communication, which can result in an uncomfortable immersive experience. The communication delay in 5G can be reduced to less than 10ms. This can afford the frame rates of 90 FPS leading to real-time 3D modeling and a comfortable immersive experience. 

The IoT network should ensure that AR/VR applications of multiple users are highly synchronized with a low latency~\cite{dhelim2022edge}. This requirement is a guarantee for supporting the users' smooth immersive experiences whenever and wherever possible. This means that, on the one hand, the network requirements are at least beyond 5G (B5G) or even 6G~\cite{alsharif2020sixth}; on the other hand, huge data monitoring and real-time computing can further boost the capacity demand in cloud computing.

\subsection{Scalable cyber worlds}
The Metaverse has been listed as one of the top five emerging trends and technologies of the next ten years. Global spending on IoT development is expected to rise from \$12 billion in 2020 to \$72.8 billion in 2024. More than 200 major brands such as Nike, Samsung, and JP Morgan, have already shifted some of their customer-oriented activities to the Metaverse. It can be envisioned that more and more IoT devices will be empowered to create and populate virtual worlds for users in social media with relative ease and low barriers to access services. As the IoT applications grow day by day, scalability is going to be a major issue with any Metaverse platform. The Metaverse must be constructed on a decentralized architecture to enhance its scalability. The user's location information and scene status in real-time interaction need to be stored offline.

Bandwidth that defines the amount of data transmitted over time is a critical requirement for the scalability of the IoT networks in the Metaverse. The bandwidth requirement is huge for most Metaverse applications, like AR/VR games. Those applications typically execute many uploading and downloading operations in video streaming, transaction, etc. The downloading and uploading demand increases as more participants join the Metaverse. Millions of users could use the Metaverse simultaneously, while the data transfer rate should remain 100 Mbps for the individual user. To maintain the quality of experience, scalable AI and communication techniques will be required. 

\subsection{Connected mobile users' experience}
Low earth orbit (LEO) satellites are developed to provide internet services to help achieve universal connectivity. Compared to a geosynchronous (GEO) satellite, the LEO satellite shortens the distance that the signal has to travel and reduces the latency. For example, the TDMA-based Iridium (by Motorola) and the CDMA-based GlobalStar are the two typical and operational LEO satellite systems. 
However, the LEO satellite system must be built with many satellites to support connected users' experience in the Metaverse. Moreover, the data packet is routed among the LEO satellites until the data can be forwarded to the end user. The satellite-to-satellite communication introduces tremendous forwarding delays.

Unlike the LEO satellite systems, in the ground high-speed data networks, the communication distance between the IoT device and the BS is short. In a cellular radio access network, the BS can tightly schedule when the IoT device sends or receives data, which increases communication efficiency. The BS can also enable the prioritization of the application according to different access capacities and latency. However, this comes at the expense of a phenomenally complex network management protocol. 

AR/VR is highly graphics-intensive and is limited by the power of the IoT device's processor. The cloud at the BS can be leveraged for graphic processing. For example, Google's Stadia gaming platform allows users to play online games, while the cloud processes their data and the user's IoT device streams the AR/VR output. The latency has to be highly reduced for the AR/VR application, where small delays of 30 ms can cause discomfort and even sickness.

\subsection{Interoperability and uniformity of virtual platforms}
Building the Metaverse has to rely on a virtual platform that allows the developers to implement IoT applications and services, such as VR/AR, autonomous driving, or E-shopping. Many Metaverse platforms are launched onto the market to replace social media in the near future, for example, Sandbox, Decentraland, Axie Infinity, Metahero, Bloktopia, etc. 

However, the lack of interoperability and uniformity between the various Metaverse platforms is one of the biggest challenges. 
Due to differences in the API standard, the current Metaverse platforms do not allow data exchange with any other platforms. This restricts the IoT applications' access in different service domains. In an ideal Metaverse, users in the physical world expect to have an uninterrupted immersive experience across multiple virtual platforms. Their physical assets can be transferred from one place to another without trouble. Since the Metaverse is expected to provide a variety of digital services, the IoT applications need to be able to seamlessly switch among the various interconnected virtual worlds.

Ensuring interoperability requires achieving interoperability between the users and the platforms, between the platforms, and between the different operating systems. One of the major advances of the Metaverse is to propose a virtual world~\cite{han2022virtual} in which the users can join and work together on different activities such as playing games, watching movies, and working. From the current situation, there is more than one company developing the Metaverse platforms~\cite{mystakidis2022metaverse}, prominent among them are Facebook, Microsoft, and Apple~\cite{9728808}. The IoT networks also need different devices to join multiple Metaverse platforms and interact with different groups. In addition, ensuring interoperability~\cite{gadekallu2022blockchain} also includes compatibility between different activities, locations, and events in the Metaverse, as well as interoperability of heterogeneous networks in the platform. 
Therefore, the future Metaverse has to be built on an interconnected virtual platform that transcends the borders of different IoT applications. There will need to be a web of public and private standards, norms, and rules for diverse IoT applications to operate across jurisdictions. 

\subsection{The barriers of the physical world}
To satisfy the ``anywhere, anytime and any participant'' requirement in the IoT-empowered Metaverse, a variety of techniques in terms of the four pillar, such as spatial computing, distributed computing, platforms/solutions for cross-cloud, have been studied. Those techniques can enhance the human interaction with the IoT devices that manipulate referents to real objects and spaces. 

Currently, the Metaverse is driven mainly by internet giants, such as Tencent, Xiaomi, Apple, and Facebook~\cite{Huang2022}. Due to interests or national policies, the virtual world created by the Metaverse is often limited by the platforms and operating systems. Once the platform is closed or inaccessible, ``anywhere, anytime and any participant'' cannot be realized. Interoperability between different operating systems, such as IOS and Android~\cite{menhaji2022systematic} users, also has many problems. To achieve seamless access, breaking down barriers between the platforms and operating systems is necessary, which is not easy to achieve. 

At present, the Metaverse highly depends on AR/VR technologies and IoT devices~\cite{lee2021all}. Most of them are difficult to adopt because the hardware is not lightweight, portable, or affordable~\cite{9757485}. Computing power can be solved by transferring the computing core to the phone or the cloud, while the battery life is difficult to break through in a short time~\cite{kim2022meta}. There is also a need to develop diverse experiences in software and hardware portability~\cite{zhu2022metaonce}. Diversified devices are not yet lightweight.

To realize the Metaverse, advanced technology and algorithms are essential~\cite{huynh2022artificial}. Firstly, the Metaverse relies on high-quality, high-performance models that can achieve the right retina display and pixel density for a realistic virtual immersion~\cite{xu2022full}. Secondly, the Metaverse needs to be powered by other technologies such as AI, AR/VR, blockchain, and web~\cite{cheng2022will} to ensure that the Metaverse can create secure, scalable, and realistic virtual worlds on a reliable and always-on platform. Thirdly, the Metaverse services need to be applied to people's daily life, where the new IoT technologies and hardwares can be validated and improved.

%=============================================================================%
%============================Section 6 Conclusion=================================%
\section{Conclusion}
\label{sec_cond}
This paper presents the IoT-inspired applications that can provide users immersive cyber-virtual experiences in healthcare, education, smart city, entertainment, real estate, and socialization. Four pillar technologies are comprehensively surveyed with their typical IoT applications in the Metaverse, including responsible AI, high-speed data communications, cost-effective MEC, and digital twins. The four pillar technologies can be leveraged to bridge the IoT applications with the Metaverse, which aims to achieve convergence of physical and cyber worlds. For each pillar technology, we focus on the motivation, significance, key technologies, and typical applications in the Metaverse. Furthermore, we extend the discussions about the critical open issues for further development and implementation in the IoT-inspired Metaverse in terms of data processing, security and privacy, real-time 3D modeling, scalable cyber worlds, connected mobile users' experience, interoperability and uniformity of virtual platforms, and barriers of the physical world.

%\section*{Acknowledgements}

\ifCLASSOPTIONcaptionsoff
  \newpage
\fi

\bibliographystyle{IEEEtran}

\bibliography{bibMeta}

% that's all folks
\end{document}